\documentclass[aps,pra,twocolumn,superscriptaddress,longbibliography,footinbib]{revtex4-2}  %
\usepackage{graphicx} 
\usepackage{float} 
\usepackage{dcolumn} 
\usepackage{bm,color}
\usepackage{amsmath,amssymb,dsfont,amstext,amsfonts}
\usepackage{extarrows}
\usepackage{xcolor}
\usepackage{wasysym}
\usepackage{mathtools}
\usepackage{bbold}
\usepackage[export]{adjustbox}
\usepackage{mathdots}
\usepackage{siunitx}
\usepackage[colorlinks=true,linkcolor=blue,urlcolor=blue,citecolor=blue]{hyperref}

\usepackage[normalem]{ulem} 
\usepackage{color}

\begin{document}

\title{A comprehensive exploration of structural and electronic properties of Molybdenum
clusters}

\author{Yao Wei}

\affiliation{Theory and Simulation of Condensed Matter (TSCM), King's College
London, Strand, London WC2R 2LS, United Kingdom}

\author{Lev Kantorovich}

\email{lev.kantorovitch@kcl.ac.uk}
\affiliation{Theory and Simulation of Condensed Matter (TSCM), King's College
London, Strand, London WC2R 2LS, United Kingdom}

\begin{abstract}
Molybdenum clusters, characterised by their unique structure and intriguing
catalytic properties, have gained significant attention in recent
years. In several existing studies density functional theory (DFT)
methods have been used to find the lowest energy Mo clusters and explore
their electronic and magnetic structure. In all cases, with the exception
of a single recent study, where a genetic algorithm was employed,
initial geometries of the clusters, prior to geometry optimisation,
were chosen using heuristic approaches based on symmetry considerations
and known structures. DFT calculations were performed using different
types of pseudopotentials, from hard to soft, and different types
of basis sets. However, no comprehensive study has yet been done in
which a DFT method with the best control on its precision would be
complemented by a reliable global minimum search method to find the
lowest energy Mo clusters. In this work, we employ a combination of
a plane wave-based DFT method and \emph{ab initio} random structure
searching (AIRSS) technique to find the lowest energy clusters of
up to 10 Mo atoms. In each case, the search has been performed for
clusters with different spin multiplicities, which enabled us to explore
their magnetic structure. The results are compared for both hard and
soft pseudopotentials stressing the importance of treating more electrons
explicitly, in agreement with some of the previous studies. For most
of the low-energy magnetic structures found, we investigate the distribution
of their spin densities, and for all low energy clusters, we confirm
their stability by calculating their phonon structure. Finally, free
energies of the Mo clusters, within the quasi-harmonic approximation,
are also calculated and discussed.
\end{abstract}
\maketitle

\section{Introduction}

Adsorption and catalysis play vital roles in the fields of physics
and chemistry. Transition metal (TM) clusters, particularly Molybdenum
(Mo) clusters, have demonstrated great potential as adsorbents for
removing heavy metal ions, organic pollutants, and radioactive nuclides
from water \cite{shao2023mixed}. Furthermore, molybdenum exhibits
catalytic properties in various processes, including hydrodesulfurization
\citep{yu2021atomic}, methanol synthesis \citep{duyar2018highly},
ammonia synthesis \citep{kojima2001molybdenum}, and redox reactions
\citep{nieter2004redox}. Additionally, molybdenum serves as an excellent
catalyst support, facilitating the preparation of metal or non-metal
nanoparticles and improving their catalytic activity and stability
\citep{miao2017molybdenum}.

Over the last decade or so, Mo clusters have been extensively studied
both theoretically \citep{zhang2004nonmetallicity,yin2022structures,aguilera2008density,ziane2017density,yin2023two,min2015study,xue2010theoretical,del2016unraveling,granja2011density,chaves2017evolution,sumer2022computational}
and experimentally \citep{ji2018linking,kumar2015graphene,feliz2019supramolecular}.
In Ref. \citep{zhang2004nonmetallicity} density functional theory
(DFT) and a plane wave basis set were used to study structural, electronic,
and magnetic properties of molybdenum clusters of up to 55 atoms including
linear, planar and 3D clusters whenever possible. A hard Mo pseudopotential
was used with only $4d^{5}5s^{1}$ electrons treated explicitly, which
resulted in clusters showing a strong tendency to dimerisation. In
\citep{aguilera2008density} a localised basis set and a hard pseudopotential
were also used confirming the dimerisation phenomenon in some of the
clusters. A localised basis set was also used in \citep{xue2010theoretical}
where an adsorption of N$_{2}$ molecule on the clusters was considered.
In all studies performed with the hard pseudopotential, the lowest
energy clusters demonstrated low symmetry (distorted triangles, pyramids,
etc.). For each number $n$ of the Mo atoms in the clusters Mo$_{n}$,
several isomers were found with close binding energies. Moreover,
the binding energy and the average bond length demonstrate an oscillating
even/odd behaviour as a function of $n$. 

In \citep{min2015study} 4s and 4p semi-core states were included
explicitly in the plane wave DFT framework proving the importance
of employing a soft pseudopotential in the description of small Mo
clusters; in particular, it was demonstrated that the dimerisation
tendency between Mo atoms is drastically reduced. This conclusion
of the dimerisation effect being an artefact of using hard core pseudopotential
on Mo atoms in Mo$_{n}$ clusters with $n=2-10$ was confirmed in
a local basis set simulations \citep{del2016unraveling} where $4s^{2}4p^{6}$
electrons on each Mo were also included in the treatment explicitly.
Some of the clusters became more symmetric; several isomers within
60 meV/atom were found for each value of $n$. Moreover, atomisation
energies demonstrated a monotonic increase with $n$. They also found
that only Mo$_{3}$, Mo$_{8}$ and Mo$_{10}$ are magnetic (triplet,
quintet and triplet), while all other lowest energy clusters were
singlet. In \citep{ziane2017density} only $4p^{6}$ electrons were
additionally included into a plane wave calculation (alongside the
$4d^{5}5s^{1}$ electrons), with the main emphasis given to studying
Mo$_{n}$S clusters. A numeric atom-centred atomic basis was used
within a scalar-relativistic approach in \citep{chaves2017evolution}
where a wide range of 3d, 4d and 5d transition-metal clusters were
studied including Mo clusters Mo$_{n}$ with $n=2-10$. It is not
clear, however, which electrons were treated explicitly in these calculations.
Average bond distances, effective coordination numbers, binding energies,
magnetic and other properties were reported. In \citep{sumer2022computational}
Mo, Pt and Mo/Pt clusters were considered using a localised basis
set; scalar-relativistic effects were included indirectly via the
Mo pseudopotential (unfortunately, no information was provided on
that pseudopotential). In this work, the stability of the structures
found was checked by calculating clusters' vibrations.

In all research mentioned so far, Mo clusters were chosen by considering
various initial geometries with different symmetry including linear,
planar and 3D geometries, from both a large set of representative
structures as well as those obtained previously; in addition, when
preparing initial geometries, distorted versions of known and/or symmetric
clusters were also tried. Only recently \citep{yin2022structures,yin2023two}
a more systematic approach based on a genetic algorithm with mutations
and mating \citep{GA_review_2009} was employed for the global minimum
energy search (for clusters with $n=2-10$) using a DFT method. In
this work a localised basis set was used; however, no information
was provided on the type of the pseudopotential employed. The authors
confirmed previously found structures and discovered a few new ones
admitting that depending on the method used the actual structures
reported in the literature may differ; the found that from $n=4$
the lowest energy structures correspond to low symmetry. A number
of properties were reported including binding energy, symmetry, average
coordinate number, second energy difference, and energy (HOMO-LUMO)
gap.

Even though in \citep{yin2022structures,yin2023two} a genetic algorithm
was used to explore the coordinate space of the clusters, a localised
basis set was used, and hence it is difficult to verify its convergence,
while a plane wave basis set can be easily extended and brought to
convergence. Also, only in a few papers the obtained clusters were
checked on their stability by calculating their vibrations. Moreover,
we did not find any discussion of the spin structure of Mo singlet
and non-singlet clusters apart from the reported value of their spin
multiplicity; the actual distribution of the spin density is not normally
discussed. As far as we are aware, in all calculations reported so
far, the van der Waals (vdW) interaction \citep{liu2012critical}
was also not accounted for.

To address these issues, we present here a systematic study of Mo
clusters with $n=2-10$ using a plane wave DFT method with both hard
and soft pseudopotentials and high energy cutoff. In all calculations,
a vdW correction to the total energy was added. For each value of
$n$, spin multiplicities $M=2S+1$ of up to 7 were considered individually
to find all clusters with the binding energies within a 1.0 eV threshold,
so that all clusters MO$_{n}$ within that threshold across all considered
spin multiplicities could be predicted. To generate initial structures
prior to their geometry optimisation, we employed an \emph{ab initio}
random structure searching technique that with high probability enabled
us to obtain the lowest energy structures for each spin multiplicity
(the `putative' global minimum). All the atomic positions of the
global minimum configurations and those close to them in energy isomers
obtained in this study are reported in Supplementary Information.
The stability of each structure is confirmed by a comprehensive vibrational
analysis, and in each spin-polarised case, the distribution of the
spin density is also given. The obtained lowest-energy structures
are compared with those reported in the literature.

The plan of the paper is as follows. In the next Section, we discuss
our computational method. In Section \ref{sec:results} results are
considered. Finally, in Section \ref{sec:Conclusions} conclusions
are drawn.

\section{Computational Methods\label{sec:Computational-Methods}}

For each number of Mo atoms $n$ in the cluster, several values of
the spin multiplicity $M=2S+1$ were considered, where $S=\frac{1}{2}N_{ud}$
is the total spin with $N_{ud}$ being the difference in the numbers
of the electrons of both directions of the spin. In this study, we
limited ourselves to the values of $M=1,3,5,7$ only (correspondingly,
$N_{ud}=0,2,4,6$). Higher values of $M$ are considered unlikely
even for clusters with large $n$. A significant set of several hundreds
of initial structures was chosen, for each value of $n$ and $M$,
using a random structure search package AIRSS (Atomic and Ionic Relaxation
from Structure Searching) \citep{pickard2006high,pickard2011ab} that
is designed to explore the configurational space in order to find
the lowest energy Mo clusters. The initial structures were selected
by defining various physical conditions such as atomic distance, symmetry,
element composition, element ratio, etc. The software performs a search
through the configurational space to identify promising atomic structures. 

Once the potential structures are identified, VASP (Vienna Ab initio
Simulation Package) \citep{kresse1996efficient} is employed to perform
structural geometry relaxation of each candidate structure and compute
its binding energies (see below). The structures obtained are then
compared with each other to select distinct structures. We used sufficiently
large unit cells in each case to prevent clusters from interacting
with their images, and correspondingly the $\Gamma$ point $\mathbf{k}-$point
sampling was only employed. This allows benefiting from a faster implementation
in VASP, specifically designed to deal with real wavefunctions. This
is particularly advantageous when dealing with a large number of structures
required for high-throughput AIRSS searches.

The local structural relaxations and total energy calculations were
performed within a DFT framework using the Perdew, Burke, and Ernzerhof
(PBE) generalised gradient approximation (GGA) functional \citep{perdew1996generalized}.
Additionally, the van der Waals (vdW) dispersion correction was introduced
employing the Grimme-3 method \citep{grimme2010consistent,grimme2011effect},
and electron-ion interactions were described using projector-augmented
wave (PAW) pseudopotentials \citep{kresse1999ultrasoft}.

Two types of pseudopotentials are compared: a hard core Mo pseudopotential
that only treats explicitly $4d^{5}5s^{1}$ valence electrons on each
Mo atom, and a soft Mo pseudopotential under which $4s^{2}4p^{6}4d^{5}5s^{1}$
electrons are considered explicitly. A plane-wave kinetic-energy cutoff
of 300 eV was used, and each relaxation calculation was considered
finished when the energy threshold of $10^{-5}$ eV was reached, while
the atomic forces were not larger than 0.01 eV/$\textrm{Å}$. To assess
the accuracy of the calculations, the optimisation of Mo$_{2}$ was
performed. The computed bond lengths (1.931 Å) and bond energy (1.973
eV/atom) closely match the experimental findings (bond length: 1.93\textasciitilde 1.94
Å and bond energy: 2.06 \textpm{} 0.38 eV/atom \citep{efremov1978electronic,hopkins1983supersonic,simard1998photoionization})
when utilising the soft Mo pseudopotential, see the comparison in
Table \ref{tab:compare_two_atoms}. However, the calculated bond lengths
(1.700 Å) and bond energy (3.104 eV/atom) do not align well with the
experimental outcomes when employing a hard pseudopotential that considers
explicitly only $4d^{5}5s^{1}$ valence electrons.

\begin{table*}[t]
\begin{centering}
\begin{tabular}{|c|c|c|}
\hline 
 & bond length (Å) & bond energy (eV/atom)\tabularnewline
\hline 
\hline 
experiment \citep{efremov1978electronic,hopkins1983supersonic,simard1998photoionization} & 1.93\textasciitilde 1.94 & 2.06 \textpm{} 0.38\tabularnewline
\hline 
this paper $4s^{2}4p^{6}4d^{5}5s^{1}$ & 1.931 & 1.973\tabularnewline
\hline 
this paper $4d^{5}5s^{1}$ & 1.700 & 3.104\tabularnewline
\hline 
\end{tabular}
\par\end{centering}
\caption{The comparison of two pseudopotemtials for a cluster with two atoms\label{tab:compare_two_atoms}}
\end{table*}

The binding energy of a Mo$_{n}$ cluster is defined in the usual
way as
\begin{equation}
E_{b}=\frac{1}{n}\left(nE_{1}-E_{n}\right)\,,\label{eq:binding-energy}
\end{equation}
where $E_{1}$ is the total energy of a single Mo atom and $E_{n}$
is the total energy of the cluster of $n$ Mo atoms.

All calculations were performed using spin-polarisation and by fixing
the number of spin-up and spin-down electrons, even if these two numbers
were the same ($M=1$). The spin density $s(\mathbf{r})=\rho_{\uparrow}(\mathbf{r})-\rho_{\downarrow}(\mathbf{r})$
is defined as a difference of the electronic densities of either of
the spins. When integrated over the whole space of the cluster, it
gives the number $N_{ud}$ of the electrons with unpaired spins. To
preview the spin density, we used VASPKIT \citep{wang2021vaspkit}
for DFT output post-processing and VESTA \citep{momma2011vesta} for
visualisation.

For each cluster obtained using AIRSS an additional geometry relaxation
was performed with more stringent relaxation criteria (the maximum
atomic force 0.01 eV/atom) prior to running the vibrational calculations.
The latter calculations were performed using finite differences (the
frozen-phonon method) as implemented in the VASP code.

To calculate the (Helmholtz) free energy of the clusters, we used
the quasi-harmonic approximation \citep{kantorovich2004quantum} with
the free energy approximated by

\begin{equation}
F=E_{DFT}+\sum_{i=1}^{3n-6}\left[\frac{\hbar\omega_{i}}{2}+k_{B}T\,\ln\left(1-e^{-\beta\hbar\omega_{i}}\right)\right]\,,\label{eq:free-energy}
\end{equation}
where $E_{DFT}$ is the DFT total energy and the second term accounts
for the contribution from atomic vibrations. There, we sum over all
vibrational frequencies $\omega_{i}$ excluding rotations and translations
(there are exactly six of these to be excluded in all cases considered
below as linear clusters have far lower binding energies than non-linear
ones \citep{yunker2011rotational} and have been excluded from the
very beginning). The first term within the square brackets is zero-point
vibrational energy, while the second term comes from taking into account
the possibility for the phonons to be excited; $k_{B}$ is the Boltzmann
constant and $T$ absolute temperature. Six frequencies appear in
these calculations as having either very small positive and/or imaginary
values and were easily eliminated; this step was also verified by
previewing the corresponding normal modes using the Interactive phonon
visualiser tool \citep{tools-phonon-dispersion} and making sure that
the eliminated modes correspond to either a translation or rotation
of the whole cluster.

\section{results\label{sec:results}}

We have used the DFT method and the AIRSS code to find the lowest
energy Mo clusters of different multiplicities with the number of
Mo atoms $n$ ranging between 2 and 10.

\begin{figure}
\begin{centering}
\includegraphics[height=6cm]{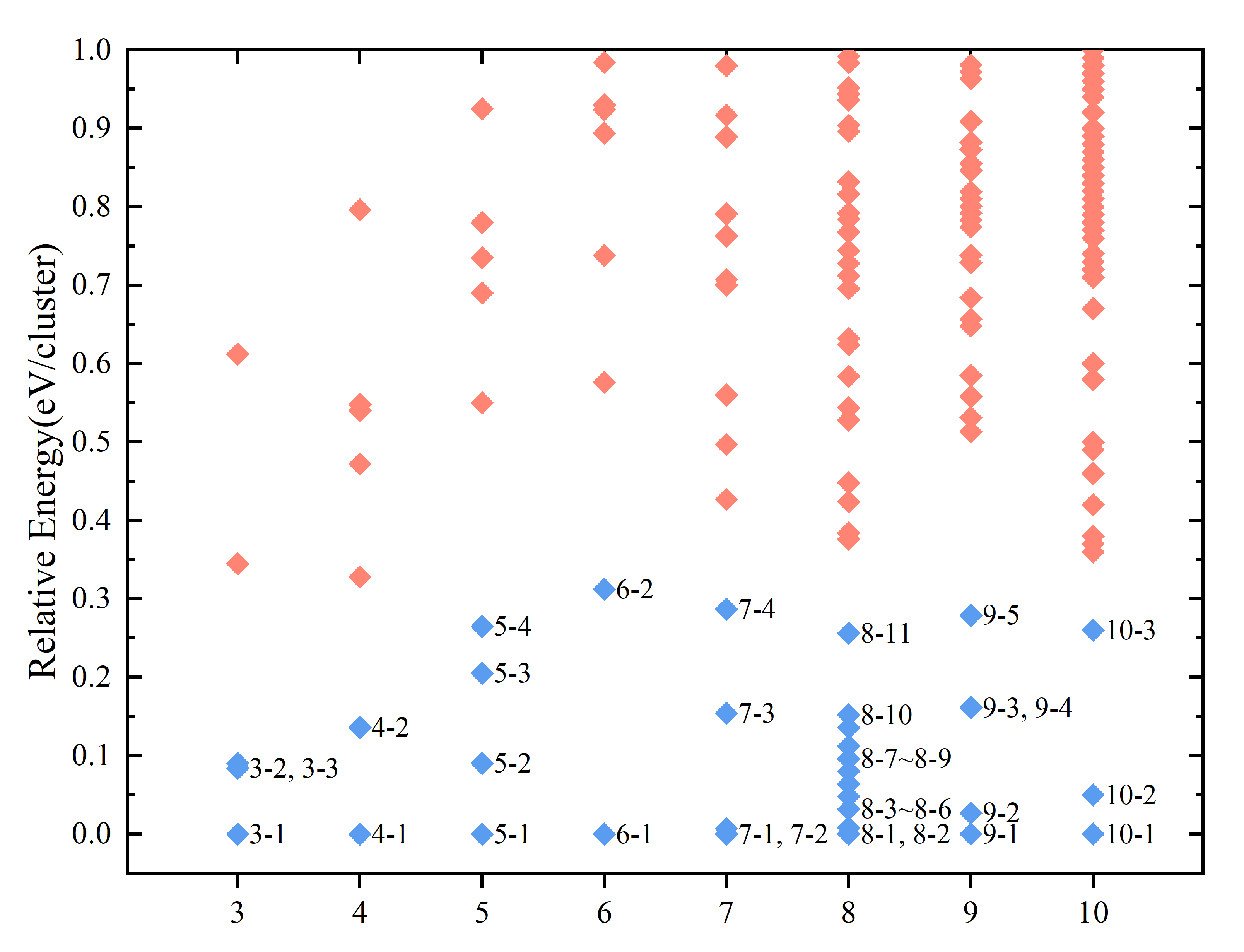}
\par\end{centering}
\caption{Energies (in eV) of DFT relaxed Mo clusters containing between $n=3$
and $n=10$ Mo atoms are shown. For each $n$ the energies are given
relative to the cluster with the lowest energy (indicated as having
0.0 energy). The clusters are numbered according to the notations
adopted in Table \ref{tab:energies_2-5}, and for most of the clusters
their point group symbol is also given. The clusters with the lowest
energies that are considered in ore detail in this work are shown
using red diamonds; others are indicated by black diamonds.\label{fig:relative_energies}}
\end{figure}

\begin{figure*}[t]
\begin{centering}
\includegraphics[width=14cm]{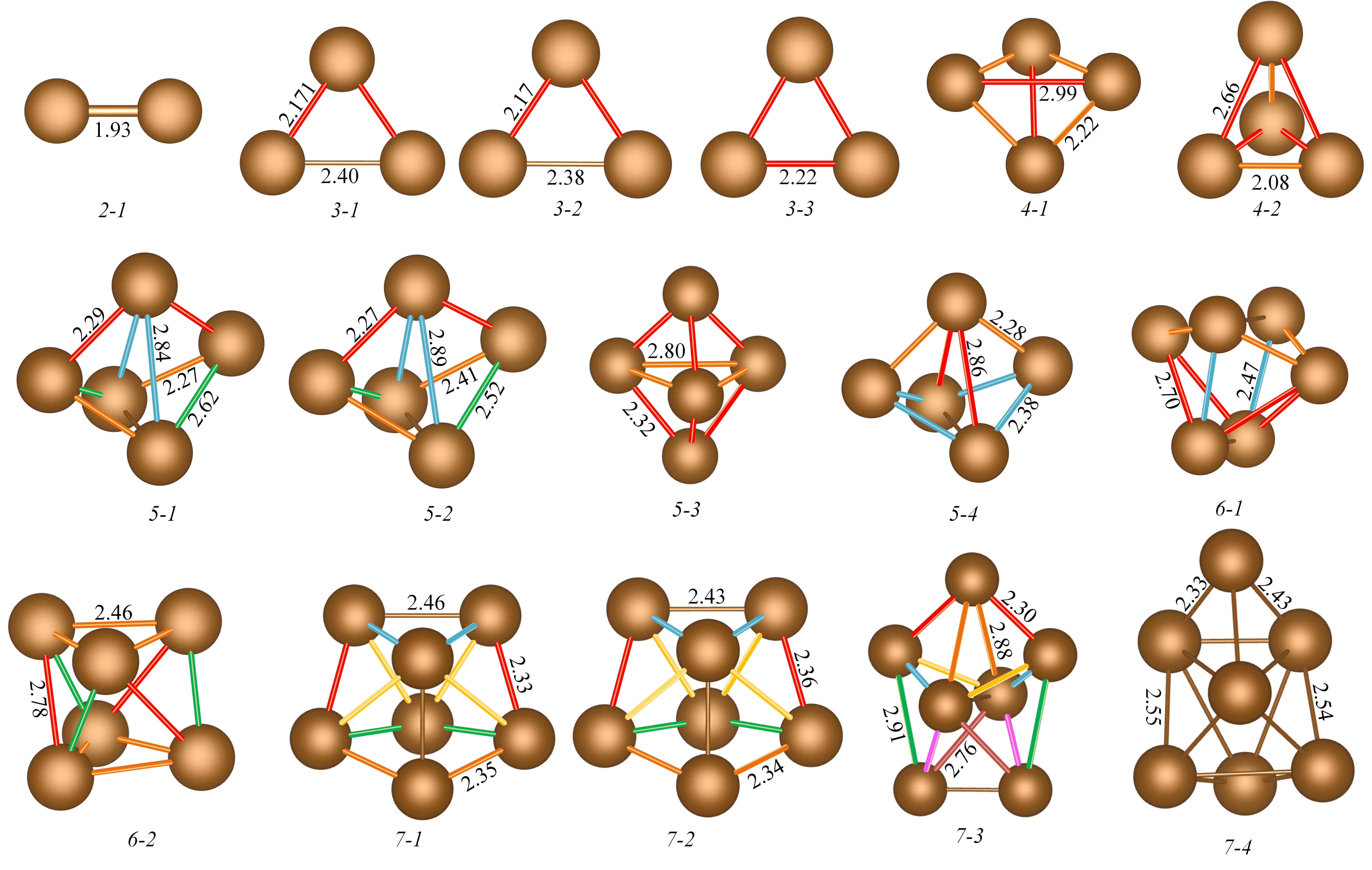}
\par\end{centering}
\caption{The lowest energy clusters for $n=2-7$ labeled as in Tables \ref{tab:energies_2-5}
and \ref{tab:energies_6-7} are shown. We also show the interatomic
distances (in $\textrm{Å}$); equal distances (within a tolerance
of about 0.01 $\textrm{Å}$) are shown, for each cluster, with the
same colour. \label{fig:geometries_2-7}}
\end{figure*}

To initiate our discussion, we first would like to show, for each
$n$, all structures we have found that are within, say, 1.0 eV in
their binding energy from the lowest energy structure. Figure \ref{fig:relative_energies}
illustrates this. The energy of the structure with the lowest energy
for any given value of $n$ is defined as 0. One can see that the
number of structures we found could be in some cases very large. Three
points can be stressed here. First of all, we see that in some cases
($n=3-7$) there is a relatively small number of clusters close in
energy to the best structure (within 1.0 eV).; in other cases ($n=8-10$)
the number of clusters within this energy range is much larger (41,
36, and 64, respectively). Secondly, for $n=3-6,10$ there is a clear
energy gap between the lowest energy structure and the next one; in
the cases of $n=7-9$ , there is a very small energy difference between
the best energy structures. Note, however, that these lowest energy
clusters may not have the same multiplicity as is clear from Tables
\ref{tab:energies_2-5}-\ref{tab:energies_9-10}. Furthermore, it
is evident that as the number of atoms in the clusters increases,
there is a gradual expansion in the variety of structure types that
have energies within 1.0 eV relative to the lowest energy structure.

These observations underscore the importance of conducting comprehensive
and systematic structure searches as in many cases there are structures
very close in energy and hence of a similar likelihood to exist and
be observed.

\begin{figure*}[t]
\begin{centering}
\includegraphics[width=14cm]{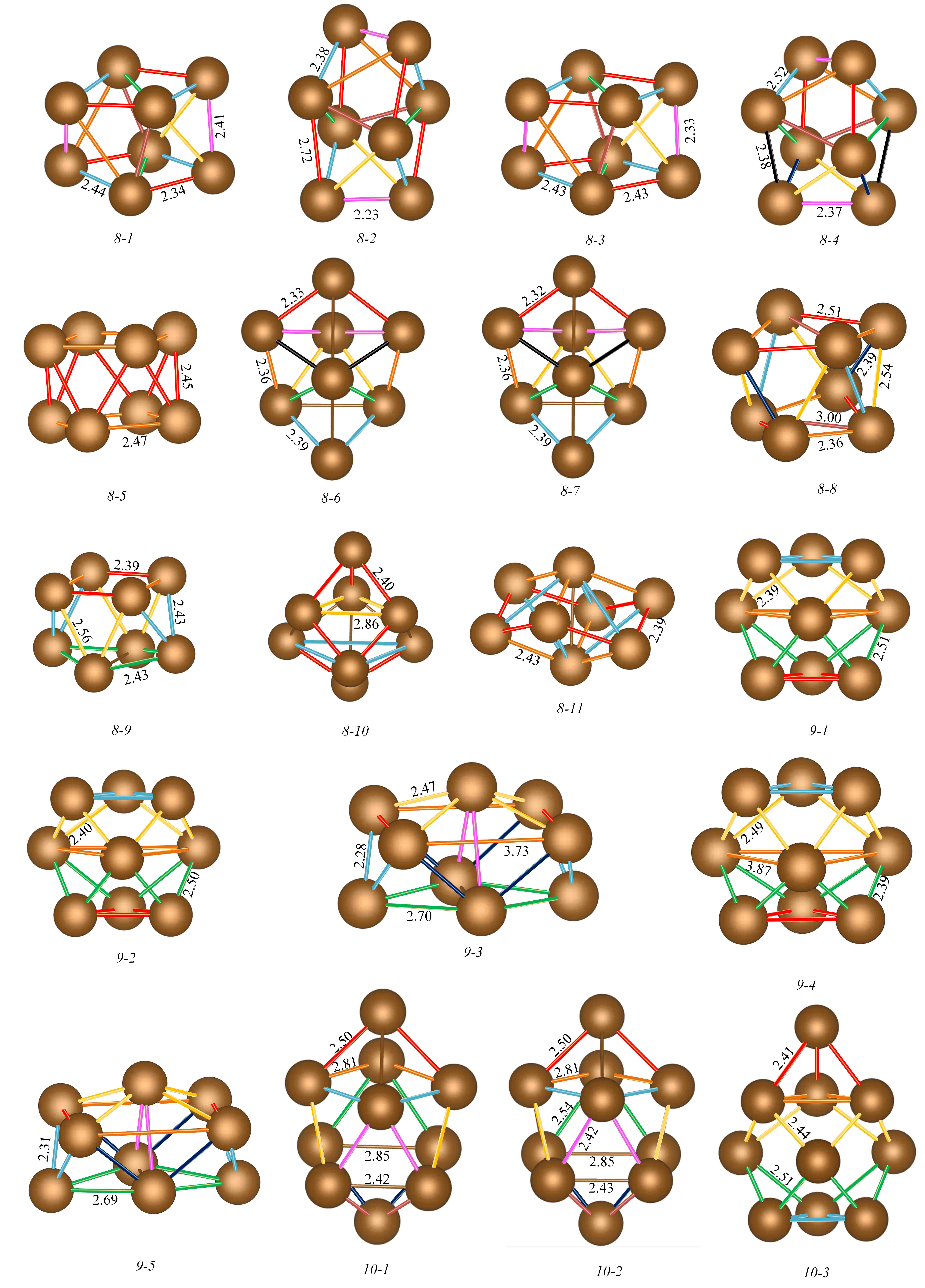}
\par\end{centering}
\caption{The lowest energy clusters for $n=8-10$ labeled as in Tables \ref{tab:energies_8}
and \ref{tab:energies_9-10}. We also show some of the interatomic
distances (in $\textrm{Å}$) with equal distances (within a tolerance
of about 0.01 $\textrm{Å}$) being shown with the same colour. \label{fig:geometries_8-10}}
\end{figure*}

Results of DFT calculations corresponding to the zero temperature
will be discussed first. Only the lowest energy structures are considered
in detail here. These are indicated in Fig. \ref{fig:relative_energies}
using blue diamonds; other structures, not discussed here in detail,
are indicated by red diamonds. The geometries of all blue structures
are given in the Supporting Information. For each value of $n$, the
lowest energy structure and a few structures with closest energies
for different multiplicities considered below in detail are shown
in Figs. \ref{fig:geometries_2-7} and \ref{fig:geometries_8-10}.
More detailed information about these structures is collected in Tables
\ref{tab:energies_2-5}-\ref{tab:energies_9-10}. Below we shall consider
each case of the number of Mo atoms $n$ separately, starting from
$n=3$.

\begin{table*}
\begin{centering}
\begin{tabular}{|c|c|c|c|c|c|c|c|c|c|c|c|}
\hline 
Cluster label & 1-1 & 2-1 & 3-1 & 3-2 & 3-3 & 4-1 & 4-2 & 5-1 & 5-2 & 5-3 & 5-4\tabularnewline
\hline 
\hline 
Multiplicity & 7 & 1 & 3 & 1 & 3 & 1 & 1 & 1 & 3 & 1 & 3\tabularnewline
\hline 
Binding energy & (-4.648) & 1.973 & 2.184 & 2.156 & 2.154 & 2.734 & 2.700 & 3.069 & 3.050 & 3.028 & 3.016\tabularnewline
\hline 
Symmetry group &  & $D_{h\infty}$ & $C_{2v}$ & $C_{2v}$ & $D_{3h}$ & $D_{2d}$ & $D_{2d}$ & $C_{2}$ & $C_{2}$ & $D_{3h}$ & $C_{2v}$\tabularnewline
\hline 
\end{tabular}
\par\end{centering}
\caption{Clusters' labels, multiplicities, binding energies (per atom, in eV),
and symmetry groups of all lowest energy clusters corresponding to
$n=2-5$. Note that for $n=1$ we give the total energy of the Mo
atom. \label{tab:energies_2-5}}
\end{table*}

\begin{table*}
\begin{centering}
\begin{tabular}{|c|c|c|c|c|c|c|}
\hline 
Cluster label & 6-1 & 6-2 & 7-1 & 7-2 & 7-3 & 7-4\tabularnewline
\hline 
\hline 
Multiplicity & 1 & 3 & 3 & 1 & 1 & 1\tabularnewline
\hline 
Binding energy & 3.423 & 3.371 & 3.562 & 3.561 & 3.537 & 3.517\tabularnewline
\hline 
Symmetry group & $C_{2v}$ & $D_{3}$ & $C_{s}$ & $C_{s}$ & $C_{2}$ & $C_{1}$\tabularnewline
\hline 
\end{tabular}
\par\end{centering}
\caption{Clusters' labels, multiplicities, binding energies (per atom, in eV),
and symmetry groups of all lowest energy clusters corresponding to
$n=6,7$.\label{tab:energies_6-7}}
\end{table*}

\subsection{Lowest energy Mo clusters}

Before we discuss in detail the Mo atom clusters we have found, a
note is in order concerning difficulties we have encountered when
trying to compare our structures with those reported in this literature.
We provide in the Supporting Information atomic positions of all high
binding energy structures and describe their symmetry in detail. However,
with a single exception of Ref \citep{chaves2017evolution}, in all
cases geometries reported in the literature are not given explicitly
via atomic positions; only pictures of the relaxed structures were
given with point groups and some interatomic distances. Hence, a precise
comparison cannot be made; hence, in doing so below we have tried
our best employing all the information available. 

\subsubsection{$n=3$}

We have found 5 isomers with the energies within 1 eV from the lowest
energy structure. Three lowest structures are shown in Fig. \ref{fig:geometries_2-7}
and their characteristics are given in Table \ref{tab:energies_2-5}.
Remarkably, the highest stability cluster labelled 3-1 is the one
with the planar geometry of an isosceles triangle having $C_{2v}$
point symmetry. The length of the two short sides of this configuration
measures 2.171Å, the longest side is 2.397 $\textrm{Å}$ and the multiplicity
$M=3$, with a binding energy of 2.184 eV/atom. Interestingly, the
structure\textit{ 3-3} corresponding to the equilateral triangle has
a lower binding energy by 0.03 eV. It has the same multiplicity and
similar bond length. Our results somewhat contradict the previous
results of Refs. \citep{aguilera2008density,min2015study,sumer2022computational},
which predicted the lowest energy structure to be also of an isosceles
triangle, but with two isosceles sides longer than the base. The studies
\citep{xue2010theoretical,yin2022structures} claimed instead that
the equilateral triangle structure is the most stable, at variance
with our results; this is most likely to do with the localised basis
set used in these studies and a hard pseudopotential (that is not
specified). However, our results generally agree with those in Ref.
\citep{chaves2017evolution,del2016unraveling,zhang2004nonmetallicity,ziane2017density}
where a similar soft pseudopotential was used.

In addition to the previously discussed structures, there is also
another structure \textit{\emph{labelled 3-2}}. The geometry of the
structure 3-2 is similar to 3-1, but it has different Mo-Mo distances
and multiplicity. The binding energy of 3-3 is by 0.002 eV lower than
that of 3-2.

The three lowest energy structures, as having almost identical energies
(the energy differences are smaller than the precision of DFT), are
all equally likely and hence all have to be accounted for in any Mo
clusters studies.

A notable observation is that the energy associated with a linear
three-atom arrangement (not shown) is elevated by approximately 1
eV per atom compared to that of the best triangular arrangement discussed
above.

\subsubsection{$n=4$}

There are seven isomers whose energies lie within 1 eV from the lowest
energy structure, two of them are analysed in detail in Fig. \ref{fig:geometries_2-7}
and Table \ref{tab:energies_2-5}. Notably, these two structures share
common traits, including the same multiplicity $M$ = 1 (and no spin
polarisation) and the (rather high) $D_{2d}$ point group symmetry.
The 4-1 structure has the lowest energy, featuring a tetrahedral-like
configuration characterised by four equally short edges measuring
2.22 Å, complemented by two extended edges of 2.99 Å, with the binding
energy of 2.734 eV/atom. On the other hand, the structure labelled
4-2 is by 0.034 eV lower in the binding energy compared to the former
structure and adopts closer to a triangular pyramid configuration
with two sides being equal to 2.66 $\textrm{Å}$ and two other sides
being shorter (2.08 $\textrm{Å}$). 

Both our geometries are of high symmetry which contradicts the result
found in \citep{yin2022structures} whose best structure has no symmetry;
their best structure also has a different multiplicity ($M=3$) to
ours. Our best structure is also quite different from the one found
in \citep{chaves2017evolution,ziane2017density,sumer2022computational}:
in \citep{chaves2017evolution,sumer2022computational} the same symmetry
$D_{2d}$ is reported and the multiplicity; however, their structure
is nearly flat; at the same time, the cluster found in \citep{ziane2017density}
has a lower $C_{2v}$ symmetry and also looks rather flat as compared
to ours. 

We exclude other structures from our discussion due to energy disparity
from the second lowest energy structure of at least 0.2 eV.

\subsubsection{$n=5$}

There are 9 isomers found in this case whose energies are within 1
eV range from the lowest energy structure. Four lowest energy configurations
are shown in Fig. \ref{fig:geometries_2-7} and their characteristics
are given in Table \ref{tab:energies_2-5}. The structure labelled
5-1 stands out with the highest binding energy of 3.069 eV/atom exhibiting
the $C_{2}$ symmetry (with the second order axis passing vertically
through the uppermost atom) and multiplicity $M$=1 (no spin polarisation).
The following next in energy structure 5-2 has the second-highest
binding energy of 3.050 eV/atom, and is of the same symmetry; however,
its multiplicity is 3 with slightly different Mo-Mo distances as indicated
in Fig. \ref{fig:geometries_2-7}. The structure labelled 5-3 presents
$D_{3h}$ point symmetry, with the third order axis passing vertically
through the top and the bottom atoms; six of nine bonds of the structure
have the same length of 2.32 Å, its configuration consists of two
regular triangular pyramids stacked together. The structure labelled
5-4 possesses a lower binding energy with multiplicity $M$= 3 and
$C_{2v}$ symmetry, the second order axis going vertically through
the upper atom. Despite the differences in the symmetry between these
four configurations, their appearances closely resemble one another.

Our best structure 5-1 agrees well with the cluster found in \citep{chaves2017evolution,sumer2022computational}
both by the multiplicity and symmetry. However, the cluster reported
in \citep{ziane2017density} has higher symmetry $C_{2v}$ and a different
multiplicity ($M=3$). Also, no symmetry for their best five atom
cluster was reported in \citep{yin2022structures} which is at variance
with our results.

\subsubsection{$n=6$}

There are 8 isomers whose energies lie within 1 eV from the lowest
energy structure, 6-1. Two lowest energy configurations are shown
in Fig. \ref{fig:geometries_2-7} and Table \ref{tab:energies_6-7}.
The binding energy of 6-1 is 3.423 eV/atom, it exhibits the $C_{2v}$
symmetry with the second order axis passing through the middle points
of the upper and lower dimers; its multiplicity is $M$= 1 (no spin
polarisation). 6-2 has higher symmetry $D_{3}$ with its upper and
lower three atoms forming parallel equilateral triangles (the lower
triangle being somewhat rotated with respect to the upper one); the
third order axis goes vertically through the centres of both triangles.
Its multiplicity is $M=3$. There is a notable energy gap of more
than 0.3 eV between the structure 6-1 and the subsequent second-lowest
energy structure 6-2.

The most stable structures of six Mo atoms reported in \citep{chaves2017evolution,ziane2017density,sumer2022computational}
have $C_{2v}$ symmetry and $M=1$ in agreement with our best structure
6-1. However, no symmetry was reported in \citep{yin2022structures},
although their multiplicity is the same as ours. 

\subsubsection{$n=7$}

We have found 14 isomers whose energies are within the 1 eV range
from the lowest energy structure. Four lowest energy configurations
are shown in Fig. \ref{fig:geometries_2-7} and Table \ref{tab:energies_6-7}.
The structure labelled 7-1 has the highest binding energy of 3.562
eV/atom exhibiting the $C_{s}$ symmetry (with the symmetry plane
passing vertically through the three atoms in the middle) with multiplicity
$M$= 3. The structure labelled 7-2 has similar symmetry and interatomic
distances; however, its multiplicity $M$= 1 is different; it exhibits
a slightly higher energy compared to the previous one. As for the
remaining two structures of the lowest energy, namely 7-3 and 7-4,
the symmetry of 7.3 is $C_{2}$ with the second order axis passing
through the upper atom vertically down through the middle points of
the other three pairs of atoms, each pair of atoms having the same
vertical coordinate, while 7-4 may seem to have $C_{s}$ symmetry
if to be judged by an eye, although a careful analysis confirms that
there is no symmetry. Structures 7-2, 7-3 and 7-4 have nearly identical
binding energies and multiplicity; however, their geometries are in
fact quite different.

We have found our structure 7-1 looking similar and of the same symmetry
$C_{s}$ as in \citep{ziane2017density,chaves2017evolution}, although
the multiplicity in \citep{chaves2017evolution} is different to ours.
The same multiplicity $M=3$ is found in \citep{yin2022structures};
however, they report no symmetry in their cluster. The best structure
found in \citep{sumer2022computational} is of the same symmetry and
multiplicity, although it looks very different. 

\begin{table*}
\begin{centering}
\begin{tabular}{|c|c|c|c|c|c|c|c|c|c|c|c|}
\hline 
Cluster label & 8-1 & 8-2 & 8-3 & 8-4 & 8-5 & 8-6 & 8-7 & 8-8 & 8-9 & 8-10 & 8-11\tabularnewline
\hline 
\hline 
Multiplicity & 5 & 1 & 3 & 3 & 1 & 3 & 1 & 1 & 1 & 1 & 1\tabularnewline
\hline 
Binding energy & 3.714 & 3.710 & 3.709 & 3.709 & 3.709 & 3.705 & 3.700 & 3.697 & 3.697 & 3.693 & 3.682\tabularnewline
\hline 
Symmetry group & $D_{2}$ & $C_{2}$ & $D_{2}$ & $C_{2}$ & $D_{4d}$ & $C_{s}$ & $C_{s}$ & $D_{2}$ & $C_{2v}$ & $T_{d}$ & $D_{3d}$\tabularnewline
\hline 
\end{tabular}
\par\end{centering}
\caption{Clusters' labels, multiplicities, binding energies (per atom, in eV),
and symmetry groups of all lowest energy clusters corresponding to
$n=8$.\label{tab:energies_8}}
\end{table*}

\begin{table*}
\begin{centering}
\begin{tabular}{|c|c|c|c|c|c|c|c|c|}
\hline 
Cluster label & 9-1 & 9-2 & 9-3 & 9-4 & 9-5 & 10-1 & 10-2 & 10-3\tabularnewline
\hline 
\hline 
Multiplicity & 3 & 1 & 1 & 5 & 3 & 3 & 1 & 1\tabularnewline
\hline 
Binding energy & 3.853 & 3.850 & 3.835 & 3.835 & 3.814 & 3.923 & 3.918 & 3.897\tabularnewline
\hline 
Symmetry group & $D_{3h}$ & $D_{3h}$ & $C_{2v}$ & $D_{3h}$ & $C_{2v}$ & $C_{s}$ & $C_{s}$ & $C_{3v}$\tabularnewline
\hline 
\end{tabular}
\par\end{centering}
\caption{Clusters' labels, multiplicities, binding energies (per atom, in eV),
and symmetry groups of all lowest energy clusters corresponding to
$n=9,10.$\label{tab:energies_9-10}}
\end{table*}

\subsubsection{$n=8$}

A total of 41 isomers fall within the 1 eV energy range of the lowest
energy structure. In Figure \ref{fig:geometries_8-10}, we present
11 lowest energy configurations, and their characteristics can be
found in Table \ref{tab:energies_8}. Interestingly, unlike clusters
composed of 2 to 7 atoms, eight-atom clusters demonstrate a significant
diversification in structural options. In this work, we focus on structures
within a narrow energy range of 0.2 electron volts, as they are of
particular interest (the next set of clusters are worse at least by
0.1 eV in their binding energies).

Configuration 8-1 has the highest binding energy of 3.714 eV/atom
exhibiting the $D_{2}$ symmetry with multiplicity $M$= 5. Eight
atoms of this cluster split into two 4-atom sets lying in two parallel
planes; one second order axis of the $D_{2}$ point group passes vertically
through the centres of the two quadrilaterals; two other perpendicular
second order axes, perpendicular to the former one, pass within the
plane located midway between the two planes of the quadrilaterals.
Compared to 8-1, structures 8-3 and 8-8 have similar symmetry, interatomic
distances and binding energies; however, all three clusters have different
multiplicities and hence electronic structures. 8-2 and 8-4 also share
the same symmetry, $C_{2}$ this time, with the 2nd order axis passing
vertically through the middle points of the pairs of upper and lower
atoms. 8-5 may look similar to 8-1 or 8-3; however, it has higher
symmetry $D_{4d}$: the two parallel quadrilaterals are perfect squares
placed symmetrically on top of each other and rotated by 45$^{\circ}$
around the vertical $C_{4}$ axis passing through the centres of the
two squares. 8-6 and 8-7 have almost identical geometry corresponding
to the group $C_{s}$; the only mirror plane passes through the four
atoms arranged vertically on the picture. 8-9 is somewhat similar
to 8-5: there are two quadrilaterals formed by the four upper and
four lower atoms; however, only the lower 4 atoms form a square, the
upper 4 atoms actually form a parallelogram reducing the symmetry
to $C_{2v}$. Cluster 8-10 has the highest $T_{d}$ symmetry with
one of the third order axis passing vertically through the upper and
lower atoms; it is only by 0.126 eV lower in binding energy than the
lowest energy structure 8-1. Finally, structure 8-11 possesses high
$D_{3d}$ symmetry with the third order axis passing vertically through
the upper and lower atoms.

Compared to available literature, our structure 8-1 looks similar
to the one reported in \citep{ziane2017density} although their structure
has different multiplicity and much higher symmetry $D_{4d}$. In
fact, their structure is comparable with ours 8-5 both in symmetry
and multiplicity. Lower symmetry clusters (and also looking rather
different) were reported in \citep{chaves2017evolution,yin2022structures};
they also have a different multiplicity to ours. One may however find
some similarities of their clusters to ours of lower energy. The cluster
found in the recent study \citep{sumer2022computational} looks similar,
but is of much lower symmetry $C_{2}$; it also has different multiplicity
to ours. 

\begin{figure*}[t]
\begin{centering}
\includegraphics[width=\textwidth]{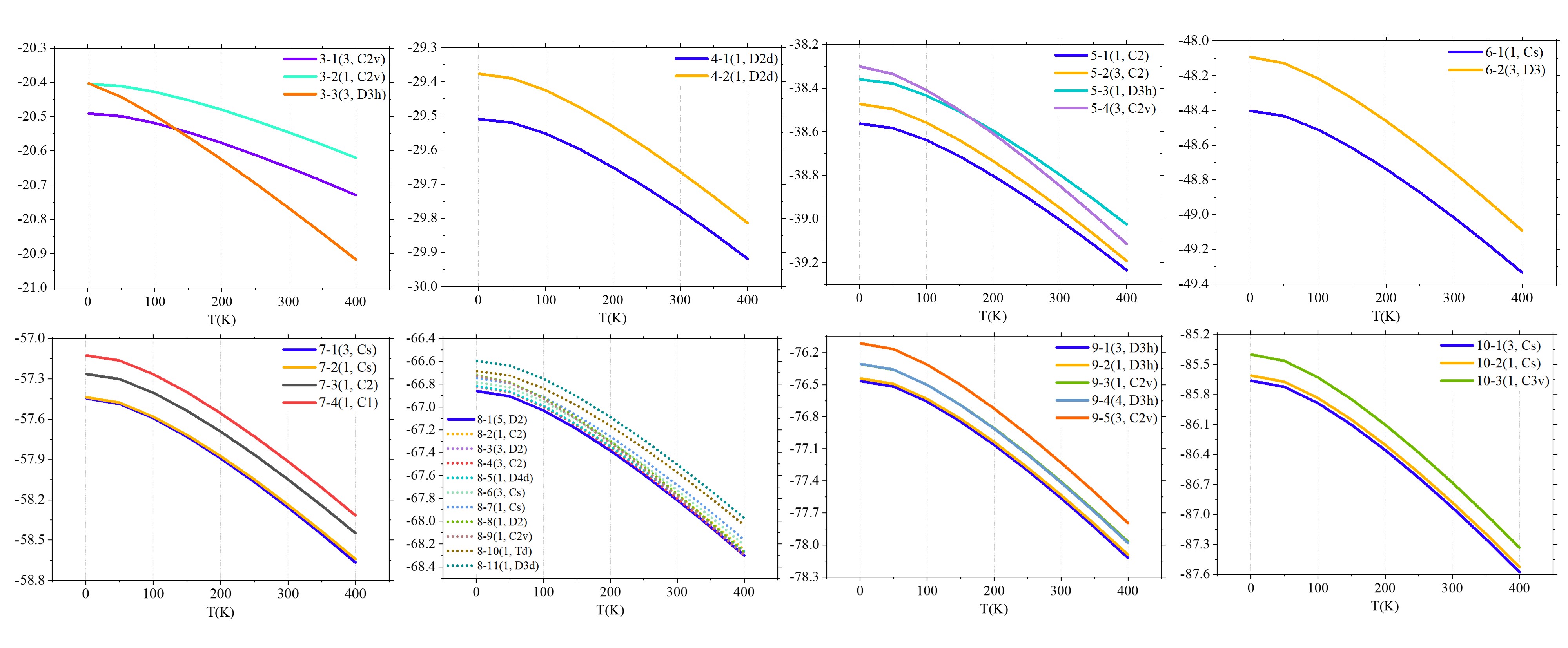}
\par\end{centering}
\caption{Free energies of DFT relaxed Mo clusters containing between $n=3$
and $n=10$ Mo atoms plotted versus temperature. \label{fig:Free-energy}}
\end{figure*}

\subsubsection{$n=9$}

We have identified 36 isomers with energies falling within the 1 eV
range of the lowest energy structure. A considerable energy gap becomes
apparent between the fifth and sixth lowest energy configurations.
In Figure \ref{fig:geometries_8-10} our investigation will be concentrated
on the five lowest energy structures, and their characteristics can
be found in Table \ref{tab:energies_9-10}.

Structures 9-1, 9-2, and 9-4 display remarkable similarities and are
very close in their energies (per atom). They have $D_{3h}$ point
group symmetry with the third order axis passing vertically through
the centres of the two equilateral triangles formed by the three upper
and three lower atoms. All three structures have different multiplicities,
however. Yet again we observe geometrically almost identical structures
and of practically the same energies (9-4 is only lower by 0.162 eV
from 9-1), but with remarkably different electronic structures. Next,
9-3 and 9-5 look similar and have the same $C_{2V}$ symmetry formed
in the following way: there are four atoms at the bottom in a form
of a rhombus; four atoms above them form a rectangular; the ninth
atom above them is placed along the vertical second order axis that
passes through the centres of both quadrilaterals. In spite of almost
identical geometry, the two clusters have different multiplicities.

Our best structure has a higher symmetry than reported in Refs. \citep{ziane2017density,yin2022structures,chaves2017evolution}:
their clusters do not have a third order axis at all. The best cluster
in \citep{chaves2017evolution} resembles our structure 9-3, both
in symmetry and multiplicity; the cluster in \citep{ziane2017density}
looks similar to ours 9-1 and has the same multiplicity, but lower
symmetry $C_{2v}$. Only $C_{s}$ symmetry cluster is reported in
\citep{yin2022structures} of the same multiplicity as ours 9-1. The
study \citep{sumer2022computational} reported the $C_{3v}$ symmetry
with the third order axis and the same multiplicity as our 9-1; however,
their structure looks very different.

\subsubsection{$n=10$}

In this case, there are 64 isomers whose energies lie within 1 eV
from the lowest energy structure, only three configurations exhibiting
the lowest energies (see Fig. \ref{fig:relative_energies}) will be
considered in more detail here; they are shown in Fig. \ref{fig:geometries_8-10}
and in Table \ref{tab:energies_9-10}. 10-1 and 10-2 share the same
$C_{s}$ symmetry and are almost identical in the binding energy (the
difference is only 0.05 eV), with the symmetry plane passing through
the four atoms forming an imaginary vertical line through the middle
of each figure. Their electronic structure is however different: if
10-1 has two unpaired electrons (triplet, $M=3$), 10-2 is in the
singlet state; however, both clusters are spin polarised (Section
\ref{subsec:Spin-density}). A more symmetrical 10-3 cluster possesses
$C_{3v}$ symmetry with the third order axis passing vertically through
the upper atom and the centre of the equilateral triangle formed by
the three atoms immediately underneath. Its energy is by 0.21 eV lower
than of 10-2; it is not spin-polarised.

The clusters proposed in Refs. \citep{ziane2017density,yin2022structures,chaves2017evolution,sumer2022computational}
have multiplicity $M=1$ and all look similar: they have four atoms
at the top, four at the bottom and two in between; neither of our
best structures shown in Fig. \ref{fig:geometries_8-10} has this
particular arrangement of atoms. Our best structure 10-1 has a very
low $C_{s}$ symmetry ($D_{2h}$ in \citep{chaves2017evolution,ziane2017density}
and no symmetry in \citep{yin2022structures}) and multiplicity $M=3$.
However, after carefully examining all the structures we have generated,
we found a structure of $D_{2h}$ symmetry and with $M=1$ that also
looks similar to the structure reported in Refs. \citep{ziane2017density,yin2022structures,chaves2017evolution};
note, however, that the binding energy of that structure is by 0.36
eV lower than that of 10-1 and hence was not included in our detailed
analysis here. 

\subsection{Free energy\label{subsec:Free-energy}}

In the preceding sections, we presented results of zero-temperature
DFT simulations based on structural searches to identify the lowest
energy configurations for clusters composed of varying numbers of
Mo atoms. Notably, our analysis revealed many clusters of close energies
(isomers), especially for large values of $n$. To shed light on the
stability of different isomers at higher temperatures, we conducted
the calculations of vibrational frequencies for all low-energy structures
(marked blue in Fig. \ref{fig:relative_energies}) and correspondingly
worked out their free energies as a function of temperature.

The vibrational frequencies of the clusters are given in the Supporting
Information for the soft pseudopotential. First of all, we note that
the frequencies are found real in all cases which proves that the
obtained geometries reported in Figs. \ref{fig:geometries_2-7} and
\ref{fig:geometries_8-10} are stable. The comparison of the vibrational
frequencies for the two types of the pseudopotentials was made only
for two clusters. We considered clusters 3-3 and 7-1, relaxed using
the hard pseudopotential and shown in Fig. S1. Their geometries are
similar to the clusters 3-1 and 7-1 obtained using the soft pseudopotential;
also, their multiplicities are the same in both cases. However, as
can be see from the comparison in Table S9, the vibrational frequencies
in both cases differ considerably. 

The calculated free energies for the temperatures of up to 400 K and
soft pseudopotentials are shown in Fig. \ref{fig:Free-energy}. The
free energy graphs yielded insightful observations. In all cases,
the free energy is going down with the temperature (as expected),
and in most cases, the order of the stability of the clusters found
at zero temperature does not change. However, there are three notable
exceptions found for the cases of $n=3$, $5$ and $8$. In the former
case, at $T=0$ cluster 3-1 is lower in energy than 3-2; however,
the order of stability of these two clusters is reversed at $T$ around
130 K and after that temperature cluster 3-2 becomes more energetically
favourable. In the case of $n=5$ the stability of the clusters 5-3
and 5-4 swapped at $T$ around 170 K, so that at higher temperatures
the cluster 5-4 is more favourable. Finally, in the case of 8 Mo atoms
clusters, the lowest energy structure 8-1 remains the most favourable
across the whole $T$ range considered; however, free energies of
structures 8-2 to 8-8 approach it rather closely, with structures
8-4, 8-5 and 8-8 having the free energy almost the same at $T=400$
K.

These intriguing observations highlight a subtle interplay in the
structural stability of some of the clusters with temperature. The
free energy calculations underscore the complex nature of cluster
stability and provide valuable contributions to the understanding
of Mo atom clusters' thermodynamic behaviour.

\subsection{Spin density\label{subsec:Spin-density}}

\begin{figure*}[t]
\begin{centering}
\includegraphics[width=14cm]{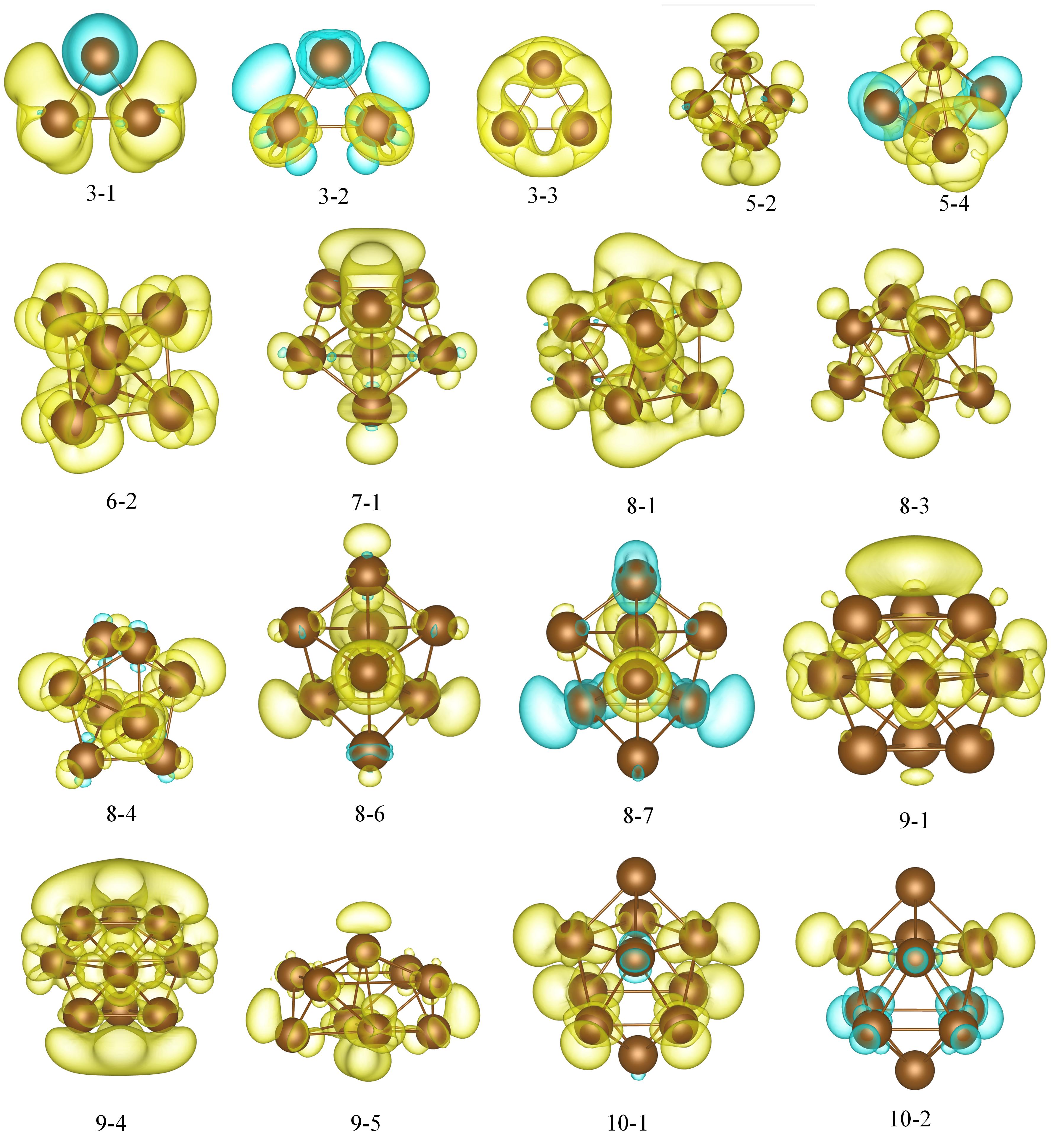}
\par\end{centering}
\caption{Spin density of DFT relaxed spin-polarised Mo clusters containing
between $n=3$ and $n=10$ Mo atoms. Constant value spin density isosurfaces
of $\pm0.003$ $\textrm{Å}^{-3}$are shown (yellow and blue correspond
to positive and negative values, respectively). The orientations of
the clusters are the same as in Figs. \ref{fig:geometries_2-7} and
\ref{fig:geometries_8-10}. \label{fig:Spin-density}}
\end{figure*}

The spin density distributions for clusters showing spin polarisation
(some of these have multiplicity $M=1$) are shown in Fig. \ref{fig:Spin-density},
and their magnetic moments are shown in Tables \ref{tab:energies_2-5}-\ref{tab:energies_9-10}. 

Note that we specifically checked that most of the clusters with $M=1$
do not have any spin polarisation even though all calculations, even
for this multiplicity, were done with spin polarisation switched on.
However, clusters 3-2, 8-7 and 10-2 do have non-zero spin density
as can be seen from the Figure: regions of both positive and negative
spin density $s(\mathbf{r})$ occupy practically equal regions of
space leading to no net spin density overall, $\int s(\mathbf{r})d\mathbf{r}=0$. 

In cluster 3-1 we have $M=3$ corresponding to two more electrons
having spin up; it is seen in the Figure that the excess of the spin-up
electrons is localised on the lower two atoms, while the upper atom
has excess of spin-down electrons. In 5-4 (also $M=3$) the excess
of the spin up density is localised mostly on the three atoms forming
a vertical plane, and there is a localisation of the spin-down density
that is distributed over two side atoms. In clusters 3-3, 5-2, 6-2,
7-1, 8-3, 8-4, 8-6 , 9-1, 9-5, and 10-1 two unpaired spin-up electrons
are distributed over the whole cluster; only small pockets of negative
spin density are seen in clusters 5-2, 7-1, 8-4, 8-6, and 10-1, these
are not expected to affect appreciably the value of their magnetic
moments. In clusters 8-1 and 9-4 we have 4 unpaired spin-up electrons
($M=5$); their density is distributed over the whole volume of the
clusters.

\subsection{Comparison with the hard pseudopotential calculations}

We have also {\small{}performed} similar calculations with a hard
pseudopotential. The results are collected in the Supporting Information.
The geometries of the clusters with lowest energies (the highest binding
energies) are shown in Fig. S1. We see, in agreement with previous
studies \citep{aguilera2008density}, that linear and planar structures
are found in this case as being of the lowest energy; in some cases,
atoms tend to dimerise (e.g., clusters 4-1, 4-4, 6-2, 9-1, and 9-3).
Both of these tendencies contradict our main results given above and
are based on calculations with soft pseudopotentials. As was also
mentioned in Section \ref{subsec:Free-energy}, vibrational frequencies
of the clusters similar in geometry and multiplicity but obtained
with two pseudopotentials, differ considerably. On the whole, we should
conclude that the results obtained with hard pseudopotentials could
be quite misleading.

\section{Conclusions\label{sec:Conclusions}}

In this work we have presented a comprehensive investigation of Mo
clusters with up to $n=$10 atoms. Using the AIRSS method \citep{pickard2006high,pickard2011ab}
we obtained a large number of clusters of lowest energies for each
value of $n$. The number of obtained clusters within 1.0 eV from
the lowest energy cluster varies with $n$, with this number quickly
growing from 5 for $n=$3 to 64 for $n=$10. In each case of $n$
for clusters with energies close to the lowest one, we have given
a detailed description of the clusters' symmetry and provided their
atomic positions and spin densities. Some of our results agree with
the clusters found by other researchers, both by their symmetry (either
exactly or very close) and multiplicity, in a few cases new clusters
were discovered, which are: 7-4, 8-1, 8-3, 8-8, 8-9, 8-10, 8-11, 9-1,
9-2, 9-4, 10-1, and 10-2. It has been found that to get reliable results,
one has to use a soft Mo pseudopotential in which $4s^{2}4p^{6}$
electrons are represented explicitly and included into the valence
shell; results obtained with a hard pseudopotential are misleading.

Generally, the low energy clusters are found to possess some symmetry
ranging from $C_{s}$ to $T_{d}$; note, however, that most of the
clusters have low symmetry like $C_{2}$ or $C_{s}$. We have also
found that in many cases there are clusters of almost identical geometries
and energies; however, their multiplicity is rather different and
hence their electronic structure. In all cases there are at least
a few clusters of almost identical energy; this is important to take
into account, e.g., when interpreting experiments \citep{loi2022oxidation}
in which monoatomic clusters are produced with a fixed number of atoms,
as there will be a distribution of clusters of different geometry
and magnetic moment in the beam. Our free energy calculations also
revealed that the order of clusters' stability may depend on temperature
in a few cases. These findings display the intricate nature of cluster
configurations and their energetic landscape. The interplay between
spin multiplicity, bond lengths, and energies adds complexity to the
understanding of these nanoscale systems.

We have also reported spin densities of all clusters with non-zero
unpaired electrons. Interestingly, in three clusters with multiplicity
$M=1$ the spin density is also non-zero with positive and negative
regions distributed equally within the cluster space.

We hope that the present study demonstrates an intriguing interplay
of spin, symmetry, and electronic properties, that would be useful
for researchers interested in physics of nanoparticles to uncover
valuable insights into the fundamental principles governing these
systems. This understanding holds promise for the development of novel
materials, catalysts, and technologies that leverage the unique characteristics
of nanoscale structures.

\section*{Acknowledgements}

Y.W. is grateful for the funding from China Scholarship Council. The
calculation is supported by UK's HEC Materials Chemistry Consortium,
which is funded by EPSRC (EP/L000202), this work used the UK Materials
and Molecular Modelling Hub for computational resources, MMM Hub,
which is partially funded by EPSRC (EP/T022213/1, EP/W032260/1 and
EP/P020194/1).


\begin{thebibliography}{0}%
\makeatletter
\providecommand \@ifxundefined [1]{%
 \@ifx{#1\undefined}
}%
\providecommand \@ifnum [1]{%
 \ifnum #1\expandafter \@firstoftwo
 \else \expandafter \@secondoftwo
 \fi
}%
\providecommand \@ifx [1]{%
 \ifx #1\expandafter \@firstoftwo
 \else \expandafter \@secondoftwo
 \fi
}%
\providecommand \natexlab [1]{#1}%
\providecommand \enquote  [1]{``#1''}%
\providecommand \bibnamefont  [1]{#1}%
\providecommand \bibfnamefont [1]{#1}%
\providecommand \citenamefont [1]{#1}%
\providecommand \href@noop [0]{\@secondoftwo}%
\providecommand \href [0]{\begingroup \@sanitize@url \@href}%
\providecommand \@href[1]{\@@startlink{#1}\@@href}%
\providecommand \@@href[1]{\endgroup#1\@@endlink}%
\providecommand \@sanitize@url [0]{\catcode `\\12\catcode `\$12\catcode `\&12\catcode `\#12\catcode `\^12\catcode `\_12\catcode `\%12\relax}%
\providecommand \@@startlink[1]{}%
\providecommand \@@endlink[0]{}%
\providecommand \url  [0]{\begingroup\@sanitize@url \@url }%
\providecommand \@url [1]{\endgroup\@href {#1}{\urlprefix }}%
\providecommand \urlprefix  [0]{URL }%
\providecommand \Eprint [0]{\href }%
\providecommand \doibase [0]{https://doi.org/}%
\providecommand \selectlanguage [0]{\@gobble}%
\providecommand \bibinfo  [0]{\@secondoftwo}%
\providecommand \bibfield  [0]{\@secondoftwo}%
\providecommand \translation [1]{[#1]}%
\providecommand \BibitemOpen [0]{}%
\providecommand \bibitemStop [0]{}%
\providecommand \bibitemNoStop [0]{.\EOS\space}%
\providecommand \EOS [0]{\spacefactor3000\relax}%
\providecommand \BibitemShut  [1]{\csname bibitem#1\endcsname}%
\let\auto@bib@innerbib\@empty
\end{thebibliography}%


\begin{thebibliography}{10}

\bibitem{shao2023mixed}
Penghui Shao, Ziwen Chang, Min Li, Xiang Lu, Wenli Jiang, Kai Zhang, Xubiao
  Luo, and Liming Yang.
\newblock Mixed-valence molybdenum oxide as a recyclable sorbent for silver
  removal and recovery from wastewater.
\newblock {\em Nature Communications}, 14(1):1365, 2023.

\bibitem{yu2021atomic}
Lei Yu, Wen-Gang Cui, Qiang Zhang, Zhuo-Fei Li, Yan Shen, and Tong-Liang Hu.
\newblock Atomic layer deposition of nano-scale molybdenum sulfide within a
  metal--organic framework for highly efficient hydrodesulfurization.
\newblock {\em Materials Advances}, 2(4):1294--1301, 2021.

\bibitem{duyar2018highly}
Melis~S Duyar, Charlie Tsai, Jonathan~L Snider, Joseph~A Singh, Alessandro
  Gallo, Jong~Suk Yoo, Andrew~J Medford, Frank Abild-Pedersen, Felix Studt,
  Jakob Kibsgaard, et~al.
\newblock A highly active molybdenum phosphide catalyst for methanol synthesis
  from co and co2.
\newblock {\em Angewandte Chemie International Edition}, 57(46):15045--15050,
  2018.

\bibitem{kojima2001molybdenum}
Ryoichi Kojima and Ken-ichi Aika.
\newblock Molybdenum nitride and carbide catalysts for ammonia synthesis.
\newblock {\em Applied Catalysis A: General}, 219(1-2):141--147, 2001.

\bibitem{nieter2004redox}
Sharon~J Nieter~Burgmayer, Dori~L Pearsall, Shannon~M Blaney, Eva~M Moore, and
  Calies Sauk-Schubert.
\newblock Redox reactions of the pyranopterin system of the molybdenum
  cofactor.
\newblock {\em JBIC Journal of Biological Inorganic Chemistry}, 9:59--66, 2004.

\bibitem{miao2017molybdenum}
Mao Miao, Jing Pan, Ting He, Ya~Yan, Bao~Yu Xia, and Xin Wang.
\newblock Molybdenum carbide-based electrocatalysts for hydrogen evolution
  reaction.
\newblock {\em Chemistry--A European Journal}, 23(46):10947--10961, 2017.

\bibitem{zhang2004nonmetallicity}
Wenqin Zhang, Xiaorong Ran, Haitao Zhao, and Lichang Wang.
\newblock The nonmetallicity of molybdenum clusters.
\newblock {\em The Journal of chemical physics}, 121(16):7717--7724, 2004.

\bibitem{yin2022structures}
Yue-Hong Yin and Jing Chen.
\newblock The structures and properties of mon (n= 2~ 15) cluster.
\newblock {\em Computational and Theoretical Chemistry}, 1212:113720, 2022.

\bibitem{aguilera2008density}
F~Aguilera-Granja, A~Vega, and LJ~Gallego.
\newblock A density-functional study of the structures, binding energies and
  magnetic moments of the clusters mon (n= 2--13), mo12fe, mo12co and mo12ni.
\newblock {\em Nanotechnology}, 19(14):145704, 2008.

\bibitem{ziane2017density}
M~Ziane, F~Amitouche, S~Bouarab, and A~Vega.
\newblock Density functional study of the structural, electronic, and magnetic
  properties of mo n and mo n s (n= 1- 10) clusters.
\newblock {\em Journal of Nanoparticle Research}, 19:1--15, 2017.

\bibitem{yin2023two}
Yue-Hong Yin and Jing Chen.
\newblock Two quasi-degenerate isomers of mo13.
\newblock {\em Journal of Cluster Science}, pages 1--11, 2023.

\bibitem{min2015study}
Byeong~June Min.
\newblock Study of the electronic and the structural properties of small
  molybdenum clusters via projector augmented wave pseudopotential
  calculations.
\newblock {\em Journal of the Korean Physical Society}, 66:209--213, 2015.

\bibitem{xue2010theoretical}
Lei Xue-Ling.
\newblock Theoretical study of small mo clusters and molecular nitrogen
  adsorption on mo clusters.
\newblock {\em Chinese Physics B}, 19(10):107103, 2010.

\bibitem{del2016unraveling}
Juli{\'a}n Del~Pl{\'a} and Reinaldo Pis~Diez.
\newblock Unraveling the apparent dimerization tendency in small mo n clusters
  with n= 3--10.
\newblock {\em The Journal of Physical Chemistry C}, 120(39):22750--22755,
  2016.

\bibitem{granja2011density}
Faustino~Aguilera Granja and Reinaldo~Pis Diez.
\newblock A density functional study of the interaction of dihydrogen with mon
  clusters (n= 2--8). adsorption and dissociation of h2 and cluster
  reconstruction after desorption.
\newblock {\em International Journal of Quantum Chemistry}, 111(12):3201--3211,
  2011.

\bibitem{chaves2017evolution}
Anderson~S Chaves, Maur{\'\i}cio~J Piotrowski, and Juarez~LF Da~Silva.
\newblock Evolution of the structural, energetic, and electronic properties of
  the 3d, 4d, and 5d transition-metal clusters (30 tm n systems for n= 2--15):
  A density functional theory investigation.
\newblock {\em Physical Chemistry Chemical Physics}, 19(23):15484--15502, 2017.

\bibitem{sumer2022computational}
Aslihan Sumer and Julius Jellinek.
\newblock Computational studies of structural, energetic, and electronic
  properties of pure pt and mo and mixed pt/mo clusters: Comparative analysis
  of characteristics and trends.
\newblock {\em The Journal of Chemical Physics}, 157(3):034301, 2022.

\bibitem{ji2018linking}
Zhe Ji, Christopher Trickett, Xiaokun Pei, and Omar~M Yaghi.
\newblock Linking molybdenum--sulfur clusters for electrocatalytic hydrogen
  evolution.
\newblock {\em Journal of the American Chemical Society}, 140(42):13618--13622,
  2018.

\bibitem{kumar2015graphene}
Subodh Kumar, Om~P Khatri, St{\'e}phane Cordier, Rabah Boukherroub, and Suman~L
  Jain.
\newblock Graphene oxide supported molybdenum cluster: first heterogenized
  homogeneous catalyst for the synthesis of dimethylcarbonate from co2 and
  methanol.
\newblock {\em Chemistry--A European Journal}, 21(8):3488--3494, 2015.

\bibitem{feliz2019supramolecular}
Marta Feliz, Pedro Atienzar, Maria Amela-Cort{\'e}s, No{\'e}e Dumait, Pierric
  Lemoine, Yann Molard, and St{\'e}phane Cordier.
\newblock Supramolecular anchoring of octahedral molybdenum clusters onto
  graphene and their synergies in photocatalytic water reduction.
\newblock {\em Inorganic Chemistry}, 58(22):15443--15454, 2019.

\bibitem{GA_review_2009}
N.~Dugan and S.~Erkoc.
\newblock Genetic algorithms in application to the geometry optimization of
  nanoparticles.
\newblock {\em Algorithms}, 2:410--428, 2009.

\bibitem{liu2012critical}
Chun-Sheng Liu, Ghanshyam Pilania, Chenchen Wang, and Ramamurthy Ramprasad.
\newblock How critical are the van der waals interactions in polymer crystals?
\newblock {\em The Journal of Physical Chemistry A}, 116(37):9347--9352, 2012.

\bibitem{pickard2006high}
Chris~J Pickard and RJ~Needs.
\newblock High-pressure phases of silane.
\newblock {\em Physical review letters}, 97(4):045504, 2006.

\bibitem{pickard2011ab}
Chris~J Pickard and RJ~Needs.
\newblock Ab initio random structure searching.
\newblock {\em Journal of Physics: Condensed Matter}, 23(5):053201, 2011.

\bibitem{kresse1996efficient}
Georg Kresse and J{\"u}rgen Furthm{\"u}ller.
\newblock Efficient iterative schemes for ab initio total-energy calculations
  using a plane-wave basis set.
\newblock {\em Physical review B}, 54(16):11169, 1996.

\bibitem{perdew1996generalized}
John~P Perdew, Kieron Burke, and Matthias Ernzerhof.
\newblock Generalized gradient approximation made simple.
\newblock {\em Physical review letters}, 77(18):3865, 1996.

\bibitem{grimme2010consistent}
Stefan Grimme, Jens Antony, Stephan Ehrlich, and Helge Krieg.
\newblock A consistent and accurate ab initio parametrization of density
  functional dispersion correction (dft-d) for the 94 elements h-pu.
\newblock {\em The Journal of chemical physics}, 132(15), 2010.

\bibitem{grimme2011effect}
Stefan Grimme, Stephan Ehrlich, and Lars Goerigk.
\newblock Effect of the damping function in dispersion corrected density
  functional theory.
\newblock {\em Journal of computational chemistry}, 32(7):1456--1465, 2011.

\bibitem{kresse1999ultrasoft}
Georg Kresse and Daniel Joubert.
\newblock From ultrasoft pseudopotentials to the projector augmented-wave
  method.
\newblock {\em Physical review b}, 59(3):1758, 1999.

\bibitem{efremov1978electronic}
Yu~M Efremov, AN~Samoilova, VB~Kozhukhovsky, and LV~Gurvich.
\newblock On the electronic spectrum of the mo2 molecule observed after flash
  photolysis of mo (co) 6.
\newblock {\em Journal of Molecular Spectroscopy}, 73(3):430--440, 1978.

\bibitem{hopkins1983supersonic}
John~B Hopkins, Patrick~RR Langridge-Smith, Michael~D Morse, and Richard~E
  Smalley.
\newblock Supersonic metal cluster beams of refractory metals: Spectral
  investigations of ultracold mo2.
\newblock {\em The Journal of Chemical Physics}, 78(4):1627--1637, 1983.

\bibitem{simard1998photoionization}
Benoit Simard, Marie-Ange Lebeault-Dorget, Adrian Marijnissen, and
  JJ~Ter~Meulen.
\newblock Photoionization spectroscopy of dichromium and dimolybdenum:
  Ionization potentials and bond energies.
\newblock {\em The Journal of chemical physics}, 108(23):9668--9674, 1998.

\bibitem{wang2021vaspkit}
Vei Wang, Nan Xu, Jin-Cheng Liu, Gang Tang, and Wen-Tong Geng.
\newblock Vaspkit: A user-friendly interface facilitating high-throughput
  computing and analysis using vasp code.
\newblock {\em Computer Physics Communications}, 267:108033, 2021.

\bibitem{momma2011vesta}
Koichi Momma and Fujio Izumi.
\newblock Vesta 3 for three-dimensional visualization of crystal, volumetric
  and morphology data.
\newblock {\em Journal of applied crystallography}, 44(6):1272--1276, 2011.

\bibitem{kantorovich2004quantum}
Lev Kantorovich.
\newblock {\em Quantum theory of the solid state: an introduction}, volume 136.
\newblock Springer Science \& Business Media, 2004.

\bibitem{yunker2011rotational}
Peter~J Yunker, Ke~Chen, Zexin Zhang, Wouter~G Ellenbroek, Andrea~J Liu, and
  Arjun~G Yodh.
\newblock Rotational and translational phonon modes in glasses composed of
  ellipsoidal particles.
\newblock {\em Physical Review E}, 83(1):011403, 2011.

\bibitem{tools-phonon-dispersion}
Elsa~Passaro Snehal~Kumbhar.
\newblock Interactive phonon visualizer tool, 2023.


\bibitem{loi2022oxidation}
Federico Loi, Monica Pozzo, Luca Sbuelz, Luca Bignardi, Paolo Lacovig, Ezequiel
  Tosi, Silvano Lizzit, Aras Kartouzian, Ueli Heiz, Dario Alf{\`e}, et~al.
\newblock Oxidation at the sub-nanoscale: oxygen adsorption on
  graphene-supported size-selected ag clusters.
\newblock {\em Journal of Materials Chemistry A}, 10(27):14594--14603, 2022.

\end{thebibliography}

\end{document}